\documentclass{article}
\usepackage{spconf,amsmath,graphicx,amssymb,color,bbm,url,bbold}
\usepackage{pgfplots, standalone}
\pgfplotsset{compat=newest}
\pgfplotsset{plot coordinates/math parser=false}
\usetikzlibrary{external}
\newlength\figureheight
\newlength\figurewidth

\newcommand{\ones}{\mathbb{1}}
\newcommand{\zeros}{\mathbb{0}}

\newcommand{\complex}{{\mbox{\bf C}}}

\newcommand{\levels}{{\mbox{\bf L}}}
\newcommand{\Tlevels}{{\mbox{\bf T}}}

\newcommand{\Tr}{\mathop{\bf tr}}
\newcommand{\diag}{\mathop{\bf diag}}

\newcommand{\var}{\mathop{\bf var}}

\newcommand{\argmin}{\mathop{\rm argmin}}

\newcommand{\vect}{\mathop{\bf vec}}

\newcommand{\matt}{\mathop{\bf mat}}

\newcommand{\bmath}{\begin{equation}}
\newcommand{\emath}{\end{equation}}

\newcommand{\bmathnt}{\begin{equation*}}
\newcommand{\emathnt}{\end{equation*}}

\newcommand{\bbmtx}{\begin{bmatrix}}
\newcommand{\ebmtx}{\end{bmatrix}}

\title{An Optimal Channel Estimation Scheme for Intelligent Reflecting Surfaces based on a Minimum Variance Unbiased Estimator}
\name{Tobias Lindstr{\o}m Jensen and Elisabeth De Carvalho}
\address{Department of Electronic Systems, Aalborg University, 9220 Aalborg, Denmark}
\begin{document}
\ninept
\maketitle
\begin{abstract}
In a wireless system with Intelligent Reflective Surfaces (IRS) containing many passive elements, we consider the problem of channel estimation. All the links from the transmitter to the receiver via each IRS elements (or groups) are estimated. As the estimation performance are dependent on the setting of the IRS, we design an optimal channel estimation scheme where the IRS elements follow an optimal series of activation patterns. The optimal design is guided by results for the minimum variance unbiased estimation. The IRS setting during the channel estimation period mimics a series of discrete Fourier transforms. We show theoretically and with simulations that the estimation variance is one order smaller compared to existing on/off methods proposed in the literature.

\end{abstract}

\section{Introduction}
\label{sec:intro}
%
For decades there's been a constant push to guarantee an increased quality of service (QoS) delivered across wireless channels. On the other hand, there's is an increased focus on energy efficiency in wireless communication that have emerged both from politics and to curb existing energy consuming technologies. An approach to address QoS and energy concerns are methods with an increased control over the propagation environment with an aim to alleviate or remove poor scattering conditions. This desired control can be achieved by introducing Intelligent Reflecting Surfaces (IRSs) \cite{Huang:2019}, where incoming signals can be reflected with passive programmable units---sometimes referred to as passive beamforming \cite{Wu:2019}. However, building this as a technology that can be deployed requires addressing a number of issues, see e.g.~\cite{Nadeem:2019} for a signal processing perspective.

One issue is channel estimation, where key problems are {\emph 1) 
as an IRS contains a large number of elements, it increases the number of links to be estimated  {\emph 2)} the IRS itself is a passive component, such that the channel can only be sensed at the receiver by sounding the channel from the transmitter, and {\emph 3)} existing IRS channel estimation schemes are more prone to errors compared to traditional non-IRS communication \cite{Nadeem:2019}. 
The existing channel estimation approaches estimate the links (or groups of links) one by one by keeping one element (or group of elements) active at each stage of the estimation process. 
%
A variant of the on/off approach is to randomly select elements ``on'' (with random phase) or ``off'' and then use sparse matrix factorization for estimation \cite{He:2019}. A different approach is to adopt the IRS to include a few active units \cite{Taha:2019}. An overview of channel estimation techniques for IRS aided communication is provided in \cite{Liang:2019, Zhao:2019}.

We consider a model with all elements passive as in \cite{Nadeem:2019, Yang:2019, Mishra:2019}. We assume no prior knowledge about the channel and employ least squares estimation since it offers a reasonably system design  \cite{Jung:2019}. Based on this model, we show that the IRS activation pattern impacts the performance of the channel estimation scheme. In particular, the on/off method \cite{Nadeem:2019, Yang:2019, Mishra:2019} is suboptimal. We design an optimal channel estimation scheme based on minimizing the Cram\'er-Rao lower bound (CRLB) under certain model constraints, including possible IRS attenuation and phase quantization. We propose a training scheme where the measurement system can be represented as orthogonal columns. 
We show that an optimal IRS activation patterns follow the rows of a the discrete Fourier transforms (DFT). Note that a DFT based training has been independently proposed in~\cite{Zheng:2019} but without optimality analysis. 
The proposed methods offers one order lower estimation variance compared to existing on/off methods. We will focus in this work on the MISO model as in \cite{Nadeem:2019, Mishra:2019}, but the ideas can also be extended to IRS OFDM models \cite{Yang:2019}.

Nomenclature: $\complex$ denotes the set of complex numbers. The operation $X = \diag(x)$ with $x\in \complex^N$ returns the matrix $X\in \complex ^{N\times N}$ with $x$ on the diagonal, i.e., such that the indices are $[X]_{i,i} = [x]_i$. With overloading $x=\diag(X)$ is the extraction of the diagonal of the matrix $X$ such that $[X]_{i,i} = [x]_i$. The operations $\vect(X)$ and $\matt(x)$ for $X\in \complex^{M\times N},x \in \complex^{MN}$ is such that $x=\vect(X)$ is a columnwise stacked version of $X$ and $X = \matt(\vect(X))$. The operation $\Tr(X)$ denotes the trace of $X$, $\angle \exp(jp) = p$ returns the angle of a complex number, $x^*$ denotes complex conjugation, $x^T$ transposition and $x^H$ is the Hermitian. Of variables we define $I_N \in \complex^{N\times N}$ to be the identity matrix, $\ones_{N} \in \complex^{N}$ a vector of ones, $\zeros_N \in \complex^N$ a vector of zeros and $E_N = \ones_{N}\ones_{N}^T$ as a matrix of ones. The operators $\odot$ and $\otimes$ denotes Hadamard (elementwise) and Kronecker product, respectively.

\section{Signal model}\label{sec:model}

\begin{figure}[tp]
  \centering
  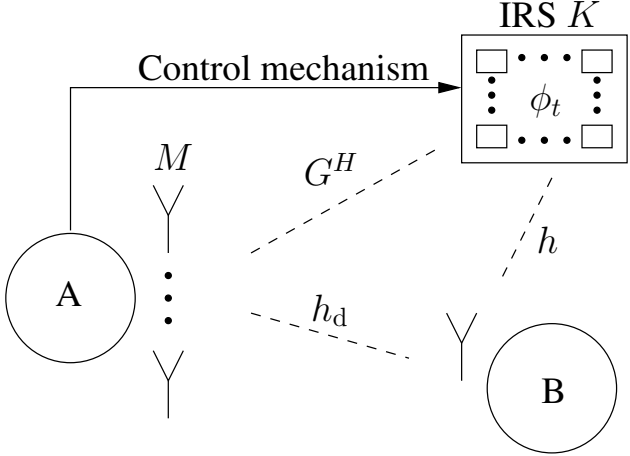
  \caption{\label{fig:model} Model of an IRS aided communication system.}
\end{figure}

We consider a MISO signal model (similar to) \cite{xianghao:2019, Wu:2019, Mishra:2019, Nadeem:2019} with a transceiver A connected to an IRS, see Fig.~\ref{fig:model}. Communication occurs using time division duplex (TDD) and we will assume reciprocity such that the downlink channel can be estimated in uplink. The node A will perform channel estimation based on data transmitted from B and set the IRS $\phi_t$ in order to control the propagation environment during training. The signal model during training step $t$ is then
\bmath
s_t = (h_{\rm d} + G^H \diag(\phi_t) h) x_t + n_t
\emath
where $s_t\in \complex^M$ is the received data, $x_t\in \complex, |x_t| = 1$ is the transmitted training symbol, $h_{\rm d}\in\complex^M $ is the direct channel between A and B, $G\in\complex^{K\times M}$ is the A to IRS channel matrix, $\phi_t = \bbmtx \phi_{t,1} & \cdots & \phi_{t,K}\ebmtx^T \in \complex^K$ is the $K$ passive elements with $\phi_{t,k} = \beta_{t,k} \exp(jp_{t,k})$, $p_{t,k} \in \levels$ are the $L=|\levels|$ quantized phase shifters and $\beta_{t,k}\in [0, 1]$ the attenuation, $h\in\complex^{K}$ is the IRS to B channel vector and $n_t \in \complex$ is additive noise. Notice that in each training step $t$ we can have different IRS settings $\phi_t$ but we assume the training can be done within channel coherence time such that $h_{\rm d}, G$ and $h$ are constant.

We can shift the order of operations and instead work with the cascaded channel $G^H\diag(h) =V = \bbmtx v_1 & \cdots & v_K \ebmtx$
\bmath \label{eq:miso_model_cascaded}
s_t = (h_{\rm d} + V \phi_t)x_t + n_t \, .
\emath
An important point is that since the IRS is passive we can not estimate $G$ and $h$ separately but only the cascaded channel $V$. We also define
\bmath
\Phi = \bbmtx 1 & \phi_{1,1} & \cdots & \phi_{1,K} \\
\vdots & \vdots & \ddots & \vdots \\
1      & \phi_{T,1} & \cdots & \phi_{T,K} \ebmtx \in \complex^{T \times K+1}, \Psi =  \Phi \otimes I_M
\emath
\bmath
s  =  \bbmtx s_1 \\ \vdots \\ s_T \ebmtx,
n  = \bbmtx n_1 \\ \vdots \\ n_T \ebmtx,
\theta =  \bbmtx h_{\rm d} \\ v_1 \\ \vdots \\ v_K \ebmtx, 
\emath
\bmath
X = \diag(\bbmtx x_1 \ones_M, \cdots, x_T \ones_M \ebmtx),  H = X \Psi .
\emath
Collecting observations $s_t$ from \eqref{eq:miso_model_cascaded} across $t=1,\ldots,T$ training periods, we observe
\begin{align}
\bbmtx s_1 \\ \vdots \\ s_T \ebmtx
 \! &= \!
\bbmtx x_1 \! \bbmtx I_M & \phi_{1,1} I_M & \cdots & \phi_{1,K} I_M \ebmtx \\
\vdots \\
x_T \! \bbmtx I_M & \phi_{T,1} I_M & \cdots & \phi_{T,K} I_M \ebmtx \ebmtx
\bbmtx h_{\rm d} \\ v_1 \\ \vdots \\ v_K \ebmtx \! + \!
\bbmtx n_1 \\ \vdots \\ n_T \ebmtx \\
&= \diag(\bbmtx x_1 \ones_M, \cdots, x_T \ones_M \ebmtx) (\Phi \otimes I_M) \theta + n \\
&= X \Psi \theta + n
\end{align}
which follows the linear measurement model
\bmath
\label{eq:linear_measurement_model}s = H \theta + n .
\emath
To reduce the number of training periods, it is possible to employ a method based on IRS blocks, where IRS units are joined together into $\bar K$ blocks with $\bar K \leq K$ \cite{Yang:2019}. In this case the signal model requires substitution $V\phi \rightarrow \bar V \bar \phi$ where $\bar V \in \complex^{M\times \bar K}, \bar \phi \in \complex^{\bar K}$ represent the block averaged channel and $\bar K$ is the number of blocks. All the subsequent analysis also applies to this case, where we will estimate $\bar V$ instead of $V$.

We will assume $n \sim \mathcal{CN}(0, \sigma^2I_{M(K+1)})$ is a circular symmetric Gaussian variable, and $T\geq K+1$, in which case the minimum variance unbiased (MVU) estimator of the channel state $\theta$ is the linear least squares estimator
\bmath
\hat \theta = \argmin \|H\theta - s\|_2^2 = (H^H H)^{-1}H^Hs
\emath
and the covariance matrix of the MVU estimator is \cite[p.~530]{Kay:1993}
\bmath
C_{\hat \theta} = \sigma^2 (H^H H)^{-1} .
\emath
The MVU estimator is efficient in that it attains the CRLB for the model \eqref{eq:linear_measurement_model}. This allows us to investigate the exact optimal statistical performance for channel estimation for this model. Notice also that the covariance matrix does not depend on the unknown channel $\theta$, and in particular not on the phase and strength of the desired signal $H\theta$, but only on the system matrix $H$ and the noise variance $\sigma^2$.

\section{Existing on/off method}
It is common in the literature to approach the channel estimation problem by switching groups of IRS elements on and off \cite{Yang:2019}, or each element \cite{Nadeem:2019, Mishra:2019}. All these approaches switch off all elements in the first phase to estimate the direct channel $h_{\rm d}$. Note that on/off corresponds to $\phi_{t, k} \in \{0, 1\}$, and in the literature $H$ is a square matrix (corresponds to $T=K+1$). Using the approach in \cite{Mishra:2019}, or the on/off ideas \cite{Yang:2019, Nadeem:2019} applied to this model, we have with the selection $K+1=T$ (up to permutation)
\bmath \label{eq:Phi_on_off}
\Phi = \bbmtx 1 & \zeros_{K}^T \\ \ones_{K} & I_{K} \ebmtx .
\emath
Note that the first column-row of $\Phi$ relates to the forced selection-estimation of the direct channel $h_{\rm d}$, respectively. The remaining columns-rows relates to the selection-estimation of $ v_1,\ldots, v_K$. The MVU estimator with the $\Phi$ selection \eqref{eq:Phi_on_off} has the covariance
\begin{align}
  C_{\hat \theta} &= \sigma^2 ((\Phi \otimes I_M)^H X^H X(\Phi \otimes I_M))^{-1} \\
  &= \sigma^2 ((\Phi^H \Phi \otimes I_M))^{-1} \\
\label{eq:Cphi}  &= \sigma^2(\Phi^H \Phi)^{-1} \otimes I_M \\
\label{eq:Cphi_on_off}  &= \sigma^2 \bbmtx 1 & -\ones_{K}^T \\ -\ones_{K} & E_K + I_K \ebmtx \otimes I_M .
\end{align}
Notice that the covariance depends on the IRS elements. The estimation variance per element (the diagonal) is
\bmath \label{eq:var_on_off}
\var([\hat h_{\rm d}]_m) = \sigma^2, \quad \var([\hat v_k]_m) = 2\sigma^2 .
\emath
A problem with this approach is that the cascaded channel is only sounded one-by-one such that the estimation variance per element is equal to $\sigma^2$, and that any error in the estimation of $h_{\rm d}$ propagates to the estimation of $v_k$. The latter is also evident from observing \cite[(11)-(12)]{Yang:2019}.

\subsection{Computational aspects}
With the on/off method where $\Phi$ is given by \eqref{eq:Phi_on_off}, the solution to the estimation problem is
\begin{align}
  \hat \theta &= (H^H H)^{-1}H^H s \\
  &= H^{-1}s \\
  &=  \vect \left( \matt(\diag(X^*)\odot s) \bbmtx 1 & -\ones_K^T \\ \zeros_{K} & I_K \ebmtx \right)\\
\label{eq:comp_on_off}  &= \bbmtx x_1^* s_1 \\ x_2^* s_2 - x_1^* s_1 \\ \vdots \\ x_T^* s_T - x_1^* s_1 \ebmtx \in \complex^{TM}\, .
\end{align}
The above can be computed with $\mathcal{O}(TM)$ operations due to the sparse $\Phi$ in \eqref{eq:Phi_on_off}.

\section{Proposed method}
We continue with the previous signal model, but instead consider another setting of $\Phi$ using the CRLB to guide the design of the channel estimation scheme. First, since the MVU estimator attains the CRLB for the linear model \eqref{eq:linear_measurement_model} we have $C_{\hat \theta} = \mathcal{I}^{-1}(\theta)$, where $\mathcal{I}(\theta)$ is the Fisher information matrix and we have the following lower bound per element
\bmath
\label{eq:C_diagonal_bound} [C_{\hat \theta}]_{i,i} = [\mathcal{I}^{-1}(\theta)]_{i,i} \geq \frac{1}{[\mathcal{I}(\theta)]_{i,i}} 
\emath
where the bound can be attained when $\mathcal{I}^{-1}(\theta)$ is diagonal (see \cite[Ex.~3.12]{Kay:1993}). From \eqref{eq:Cphi}, and without other modifications, we identify that we can attain the bound \eqref{eq:C_diagonal_bound} if $\Phi^H\Phi = \diag(d)$\footnote{It is interesting to note that it is not uncommon that design choices influences the CRLB, see e.g.~\cite{Djuric:1990, Jensen:2017} where the time index is important for chirp and harmonic chirp models.}. Furthermore, say that we will try to achieve the same variance for all unknowns $\theta$. This implies that $\Phi$ has equally scaled orthogonal columns $\Phi^H\Phi = \alpha I_{K+1}$ (or for $T=K+1$ that $\sqrt{\alpha} \Phi$ is orthogonal). In this setting, minimizing the variance of the estimate is equivalent to maximizing $\alpha$ with the constraints that the first column is $\ones_{T}$ (corresponding to the direct channel $h_{\rm d}$ that the IRS cannot control) and the rest of the elements follows the model $\phi_{t,k} = \beta_{t,k} \exp(jp_{t,k})$, with phase quantization levels $p_{t,k}\in \levels$ and attenuation $\beta_{t,k}\in [0, 1]$. Under these design constraints, an optimal training scheme for the IRS $\Phi$ is the solution to the optimization problem
\begin{equation} \label{eq:optimal}
\begin{array}{ll}
	\displaystyle \text{maximize} & \alpha \\
	\text{subject to} & \phi_{t,k} = \beta_{t,k} \exp(jp_{t,k}),  t=1,\ldots,T; k=1,\ldots,K  \\
        & \beta_{t,k} \in [0, 1],  t=1,\ldots,T; k=1,\ldots,K  \\
        & p_{t,k} \in \levels, t=1,\ldots,T; k=1,\ldots,K \\
        & [\Phi]_{t, k} = \phi_{t,k}, t=1,\ldots,T, k =2,\ldots,K+1 \\
        & [\Phi]_{t, 1} = 1, t=1,\ldots,T \\
        & \Phi^H\Phi = \alpha I_{K+1} \\
        & \Phi\in\complex^{T \times K+1}. \\
\end{array}
\end{equation}
In general, this problem can be difficult to solve, and as we will show in particular due to the possible phase quantization levels $\levels$. But a solution to \eqref{eq:optimal} can be found for a particular choice of quantization, that is if $\{0, 2\pi/T,\cdots, 2\pi(T-1)/T\} = \Tlevels \subseteq \levels$. To find this solution, first observe that the objective has the following upper bound
\begin{align}
  \alpha &= \frac{1}{K+1}\Tr(\Phi^H\Phi) \\
  & = \frac{1}{K+1}\sum_{t=1}^T \sum_{k=1}^{K+1} |\phi_{t,k}|^2 \\
    & \leq \frac{(K+1)T}{K+1} \\
 \label{eq:bound} &= T .
\end{align}
Now, let $F_{T, K+1}\in\complex^{T\times K+1}$ be the $K+1$ leading columns of a $T\times T$ DFT matrix $[F_{T, K+1}]_{t,k} = \exp(-j2\pi(t-1)(k-1)/T)$. Notice that $F_{T, K+1}$ can contain no more than $T$ unique values around the unit circle, i.e., $\angle [F_{T, K+1}]_{t,k} \in \Tlevels$. The choice $\Phi = F_{T, K+1}$ satisfies the design constraints in \eqref{eq:optimal} under the assumption that $\Tlevels \subseteq \levels$ and the upper bound \eqref{eq:bound} can be attained 
\bmath
\Phi^H \Phi = F_{T, K+1}^H F_{T, K+1} = T I_{K+1} \qquad (\alpha=T) \ . 
\emath
Thus $\Phi = F_{T, K+1}$ a solution to \eqref{eq:optimal} corresponding to having all attenuation $\beta_{t,k} =1$. In this case the IRS pattern during training will mimic the DFT matrix and is an optimal scheme under the given design constraints. The estimation covariance is then
\begin{align}
  C_{\hat \theta} &= \sigma^2(\Phi^H \Phi)^{-1} \otimes I_M \\
  &= \sigma^2(F_{T, K+1}^H F_{T, K+1})^{-1} \otimes I_M \\
  &= \frac{\sigma^2}{T} I_{M(K+1)}
\end{align}
diagonal as per design and the estimation variance per element (the diagonal) is
\bmath \label{eq:var_proposed}
\var([\hat h_{\rm d}]_m) = \var([\hat v_k]_m) = \frac{\sigma^2}{T}.
\emath
Notice that from \eqref{eq:var_proposed}, we observe that the proposed method offers one order of magnitude lower estimation variance as dictated by the factor $1/T$ compared to \eqref{eq:var_on_off}. Furthermore, with this approach we can increase the number of training symbols $T \geq K+1$ to decrease the variance of the estimate (the overcomplete case) and improve the estimation accuracy for all unknown $\theta = \bbmtx h_d^T & v_1^T & \cdots & v_K^T \ebmtx^T$.

Another optimal choice is $\Phi = P_1 F_{T,K+1} P_2$ where $P_1, P_2 \in \{0,1\}^{T\times T}$ are permutation matrices $P_1^T P_1 = P_2^T P_2 = I_T$ and $[P_2]_{1,1} = 1$ (to ensure $[\Phi]_{t,1} = 1, \forall t$). 

\subsection{Computational aspects}
With the proposed method, the solution to the least squares estimation problem is
\begin{align}
  \hat \theta &= (H^H H)^{-1} H^H s \\
  &= \frac{1}{T} (\Phi \otimes I_M)^H X^H s \\
  &= \frac{1}{T} \vect( \matt(\diag(X^*) \odot s) F_{T, K+1}^*)
\end{align}
which is no more than the application of $M$ inverse Fast Fourier Transform along the rows of $\matt(\diag(X^*) \odot s) \in \complex^{M \times T}$ followed by selection of the first $K+1$ columns of the result. So, the least squares solution is possible in $\mathcal{O}(M T \log T)$ operations, a factor of $\log(T)$ higher than \eqref{eq:comp_on_off} which is due to the introduction of dense linear algebra via the choice $\Phi=F_{T,K+1}$.

\begin{figure}
  \centering
  \setlength{\figureheight}{1.35\figureheight} 
  \includegraphics[width=0.84\linewidth]{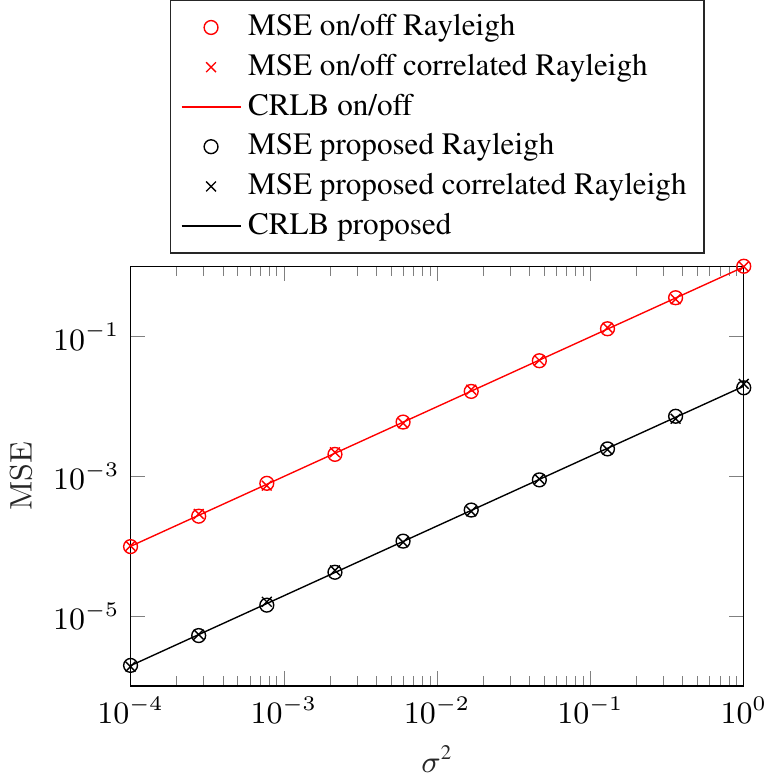}
\vspace{-0.4cm}
\caption{\label{fig:figure1_hd} Measured MSE and the estimation variance of $[h_{{\rm d}}]_1$ versus $\sigma^2$. $K = 50$, $M = 10$, $T = K + 1$.}
\vspace{-0.2cm}
\end{figure}

\section{Simulations}
To verify our claims, we present Monte Carlo simulations with $R=1000$ repetitions (independent channel and noise realizations). We measure the mean squared error (MSE) of the error $e = \theta - \hat \theta$, which on average should be the same as the variance of $e$ since the estimators are unbiased. The channel is modeled as Rayleigh fading and correlated Rayleigh fading with a correlation matrix of the form $[R]_{i,j}=r^{|i-j|}, r=0.95$. This is done to highlight that the estimator is independent of the channel, and we should observe the same estimation accuracy for both channel types. For these simulations we only consider the case $T=K+1$ as this gives a parameter setup where it is possible to compare the on/off methods as presented in the literature and the proposed method.

In Fig.~\ref{fig:figure1_hd}--\ref{fig:figure1_v} we observe an agreement between the measured MSE and the achievable CRLBs in \eqref{eq:var_on_off} and \eqref{eq:var_proposed} as expected versus noise variance $\sigma^2$ for both the standard on/off approach and the proposed method. We also observe that the estimation accuracy for both Rayleigh and correlated Rayleigh match the CRLBs. Notice that the variance of the on/off method for $v_{k}$ is twice as high as for $h_{{\rm d}}$ as also predicted in \eqref{eq:var_on_off}, and that the proposed method offers an order $T$ better estimation accuracy. Plotting the same versus $K$ in Fig.~\ref{fig:figure2_hd}--\ref{fig:figure2_v} we observe again an agreement between the measured MSE and the estimation variance. Notice the important relationship, that the estimation error of the proposed methods decreases as a function of $T = K+1$,  whereas the MSE remains constant for the on/off method. This particular relation is important, because the number of IRS passive units $K$ is often large but the estimation variance per unit $k=1,\ldots,K$ is decreasing as a function of $K$---if we still ensure $T\geq K+1$.

\section{Discussion}
In this paper we showed a optimal channel estimation scheme where the analysis was aided by known bounds for the case of linear least squares. The presented approach could offer new ideas in the design for more complicated methods and models, and adding additional assumptions in future work, or schemes aimed at shortening the training period (underdetermined estimation).

\begin{figure}
  \centering
    \includegraphics[width=0.84\linewidth]{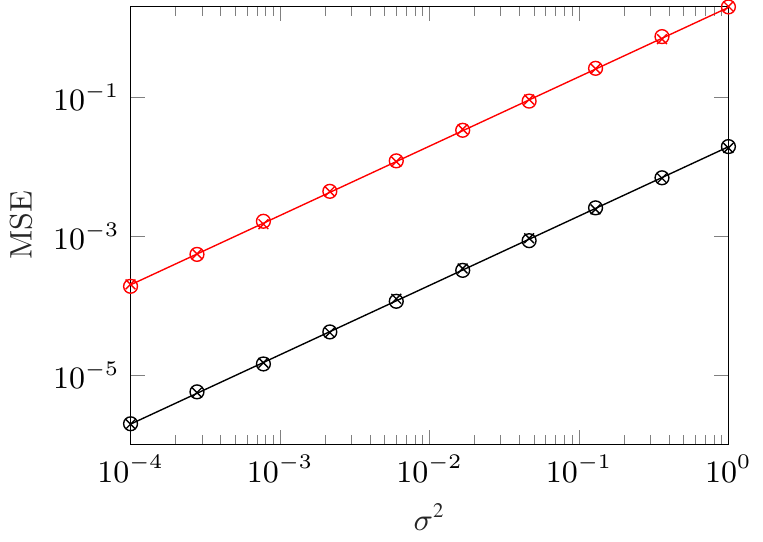}
\vspace{-0.4cm}
\caption{\label{fig:figure1_v} Measured MSE and the estimation variance of $[v_{1}]_1$ versus $\sigma^2$. $K = 50$, $M = 10$, $T = K + 1$.}
\end{figure}

\begin{figure}
  \centering
    \includegraphics[width=0.84\linewidth]{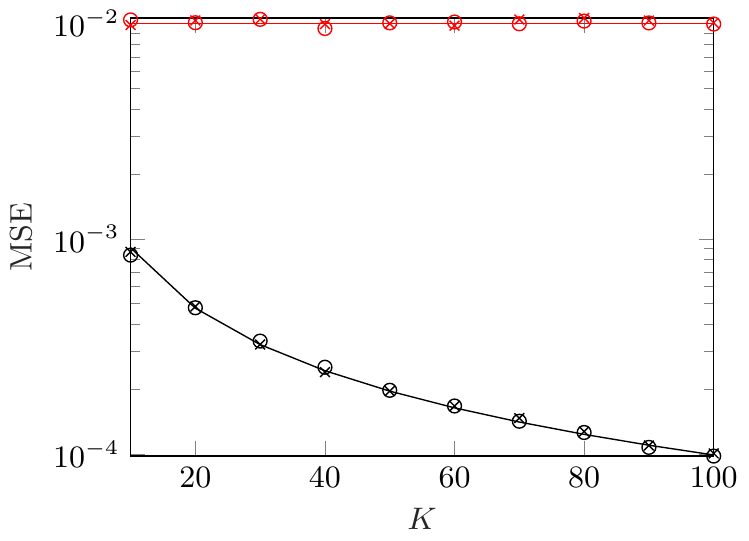}
\vspace{-0.4cm}
\caption{\label{fig:figure2_hd} Measured MSE and the estimation variance of $[h_{{\rm d}}]_1$ versus $K$. $\sigma^2 = 1\cdot 10^{-2}$, $M = 10$, $T = K + 1$.}
\end{figure}
\begin{figure}
  \centering
    \includegraphics[width=0.84\linewidth]{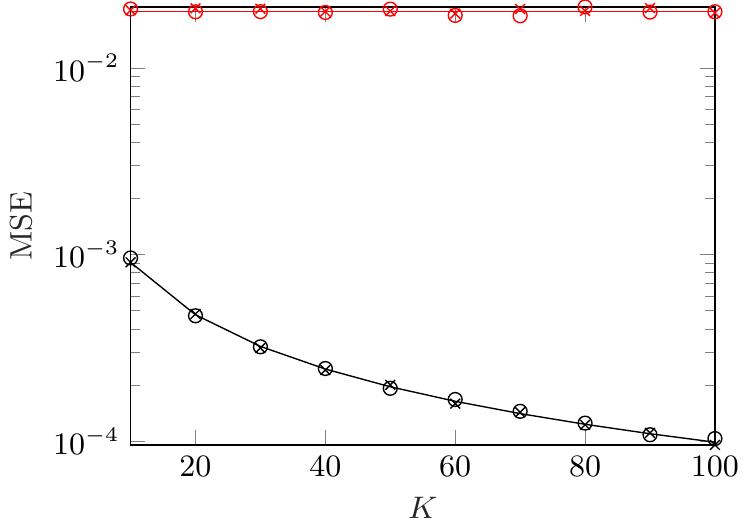}
\vspace{-0.4cm}
\caption{\label{fig:figure2_v} Measured MSE and the estimation variance of $[v_{1}]_1$ versus $K$. $\sigma^2= 1\cdot 10^{-2}$, $M = 10$, $T = K + 1$.}
\end{figure}


\newpage
\bibliographystyle{IEEEbib}
\bibliography{refs}

\end{document}